\documentclass[aps,reprint,floatfix,superscriptaddress]{revtex4-1}
\usepackage[english]{babel}
\usepackage[utf8]{inputenc}
\usepackage[dvipsnames]{xcolor}
\usepackage{geometry}
\usepackage{babel}
\usepackage{amsmath,amsfonts}
\usepackage{graphicx}
\usepackage{hyperref}

\providecommand{\tabularnewline}{\\}

\begin{document}

\newcommand{\eg}{{\emph{e.g.}\ }}
\newcommand{\ie}{{\emph{i.e.}\ }}
\newcommand{\FIXME}[1]{{\bf FIXME: #1}}

\newcommand{\HSYK}{H_{\text{SYK}}}
\newcommand{\HbSYK}{H_{\text{b-SYK}}}
\newcommand{\HamI}{\mathcal{H}_{\text{I}}}

\renewcommand{\i}{\imath}
\newcommand{\up}{\uparrow}
\newcommand{\down}{\downarrow}
\newcommand\ket[1]{\left|#1\right\rangle }
\newcommand\bra[1]{\left\langle #1\right|}
\newcommand\braket[2]{\left\langle #1\middle|#2\right\rangle }
\newcommand\ketbra[2]{\left|#1\vphantom{#2}\right\rangle \left\langle \vphantom{#1}#2\right|}
\newcommand\braOket[3]{\left\langle #1\middle|#2\middle|#3\right\rangle }

\newcommand{\rSYK}{\langle r_{\mathrm{SYK}} \rangle}
\newcommand{\rI}{\langle r_{\mathrm{I}} \rangle}
\newcommand{\rbSYK}{\langle r_{\mathrm{b-SYK}} \rangle}
\newcommand{\rEXP}{\langle r \rangle}

\title{Sachdev-Ye-Kitaev type physics in the strained Kitaev honeycomb model}

\author{Mikael Fremling}

\author{Lars Fritz}

\affiliation{Institute for Theoretical Physics and Center for Extreme Matter and Emergent Phenomena,
Utrecht University, Princetonplein 5, 3584 CC Utrecht, The Netherlands}

\begin{abstract}
In this work, we investigate whether the Kitaev honeycomb model can serve as a starting point to realize the intriguing physics of the Sachdev-Ye-Kitaev (SYK) model.
The starting point is to strain the system, which leads to flat bands reminiscent of Landau levels, thereby quenching the kinetic energy.
The presence of weak residual perturbations, such as Heisenberg interactions and the $\gamma$-term,
creates effective interactions between the Majorana modes when projected into the flux-free sector.
We assume the resulting interactions to be effectively random. This leads to a bipartite Sachdev-Ye-Kitaev model (b-SYK) with very similar properties as the SYK model.
We also hypothesize under which conditions one would expect the standard SYK model in such a setup.

\end{abstract}

\maketitle

\section{Introduction}
Some of the most important models and concepts in theoretical physics unite both high- and low-energy physics: Landau theory, renormalization group, the Higgs mechanism,
topological Chern-Simons fields theories, and Ising type models. Recently,
the Sachdev-Ye-Kitaev (SYK) model has been added to this illustrious list~\cite{Sachdev1993,Kitaev2015}. 
Its Hamiltonian
\begin{eqnarray}\label{eq_SYK}
H_{\rm{SYK}}=\sum_{i,j,k,l}J_{ijkl}\gamma_i \gamma_j \gamma_k \gamma_l\,, 
\end{eqnarray}
describes $N$ localized Majorana fermions $\gamma_i$ with $i=1,...,N$,
interacting via a random all-to-all interaction $J_{ijkl}$.
 It is usually assumed to be Gaussian with mean $\langle J_{ijkl} \rangle=0$ and variance 
 \begin{eqnarray}\label{eq_gaussdistr}
 \langle J_{ijkl} J_{i^{\prime}j^{\prime}k^{\prime}l^{\prime}}\rangle=\frac{6J^2}{N^3}\delta_{i,i^{\prime}}\delta_{j,j^{\prime}}\delta_{k,k^{\prime}}\delta_{l,l^{\prime}}\;.
 \end{eqnarray}

This model has a number of fascinating properties: it is a strongly coupled quantum many-body system that is chaotic\cite{Gu2017,Berkooz2017,Hosur2016},
nearly conformally invariant, exactly solvable in the infrared of the large-$N$ limit~\cite{Rosenhaus2019},
and a fast scrambler of quantum information having maximal Lyaponov exponents~\cite{Maldacena2016}.
It is believed to describe the essential physics of two dimensional gravity,
black holes, as well as non-Fermi liquids~\cite{Sachdev2015,Song2017}.
Further intriguing properties include unusual spectral properties~\cite{You2017,Polchinski2016,GarciaGarcia2016,cao2020towards},
patterns of entanglement~\cite{Liu2018,Huang2019},
the presence of supersymmetry~\cite{Fu2017,Behrends2020},
and unusual quantum phase transitions~\cite{Banerjee2017,Bi2017,LantagneHurtubise2018}.

To date, there is as large body of experimental proposals targeting the realization of either fermionic or Majorana versions of the SYK model.
These include realizations in superconductors~\cite{Pikulin2017},
Majorana wires~\cite{Chew2017}, graphene~\cite{Chen2018}, 
quantum simulators~\cite{Luo2019}, and optical lattices~\cite{Wei2021}.

The work presented here is inspired by Ref.~\onlinecite{Chen2018}, which investigated the realization of the fermionic SYK model using graphene.
In their work, the idea is to subject a flake of graphene to a magnetic field which leads to Landau levels,
thereby quenching the kinetic energy (this effectively localizes the electrons to within the cyclotron radius).
For a generic filling, this is a highly degenerate situation, and one has to consider the effect of Coulomb interactions and disorder to split it. 
They managed to show that the combined effect of localization within the cyclotron radius, Coulomb interaction,
and disorder at the boundary of the flake leads to an effectively zero-dimensional problem of electrons interacting with a random interaction strength.
While this leads to an effective model akin to the SYK model, the degrees of freedom of the model were fermionic and not of Majorana character.

{\it Main result:} In this work, we pursue a realization of the Majorana formulation of the SYK model.
In doing so, we find a variant henceforth referred to as the b-SYK (bipartite Sachdev-Ye-Kitaev) model.
A detailed study of its properties appears in a parallel work, Ref.~\cite{Fremling2021},
but the most important results are summarized here in a self-contained fashion.

{\it The recipe:} We start from the honeycomb Kitaev model (KHM) and achieve flat Majorana bands by applying strain~\cite{Rachel2016} effectively quenching the kinetic energy. 
The necessary interactions between the effective Majorana degrees of freedom come from generically present perturbations  in honeycomb Kitaev systems
\cite{Jackeli2009,Takagi2019},
namely Heisenberg interactions and the so-called $\gamma$-term~\cite{Rau2014,Katukuri2014,Yamaji2014}.
We model the interactions as effectively random for reasons explained later.
{\it The b-SYK model:}
We find a variant of the SYK model that we call b-SYK.
Its Hamiltonian reads
\begin{eqnarray}\label{eq_bSYK}
H_{\rm{b-SYK}}=\sum_{i,j}^{N_A} \sum_{k,l}^{N_B}J_{ijkl}\gamma_i^A \gamma_j^A \gamma_k^B \gamma_l^B \,,
\end{eqnarray}
with two sets of Majorana fermions, $A$ and $B$, consisting of $N_A$ and $N_B$ Majorana fermions, respectively. 
The effective coupling in the model is random with $\langle J_{ijkl} \rangle=0$ and 
 \begin{eqnarray}\label{eq_bgaussdistr}
 \langle J_{ijkl} J_{i^{\prime}j^{\prime}k^{\prime}l^{\prime}}\rangle=\frac{6J^2}{\sqrt{N_A N_B}^3}\delta_{i,i^{\prime}}\delta_{j,j^{\prime}}\delta_{k,k^{\prime}}\delta_{l,l^{\prime}}\;.
 \end{eqnarray}
 The model itself is the subject of a more in-depth study in a parallel paper~\cite{Fremling2021}:
 It has tuneable scaling dimensions for the Majorana fermion operators in the conformal limit,
 and it has level statistics that is different from the SYK model, which is discussed in this paper.

{\it Outline of the paper:}
In Sec.~\ref{sec:background} we introduce the main ingredients as well as the main idea of the paper.
Concretely, we discuss the solution of the KHM in terms of Majorana fermions in Sec.~\ref{sec:KHM} and proceed with the role of strain in Sec.~\ref{sec:Strain}.
We also investigate the role of the flux gap in the strained system in Sec.~\ref{sec:Fluxgap}.
In Sec.~\ref{sec:Interaction} we lay out a route leading to an effective SYK type model emerging from perturbations on top of the strained KHM.
Sec.~\ref{sec:Elems} discusses the effective microscopic Hamiltonian and to which extent the couplings fulfill the requirement of being random and realizing SYK type physics.
We conclude with a summary, a critical discussion, and an outlook in Sec.~\ref{sec:summary}.

\section{Background}\label{sec:background}
\begin{figure}
  \includegraphics[width=0.3\textwidth]{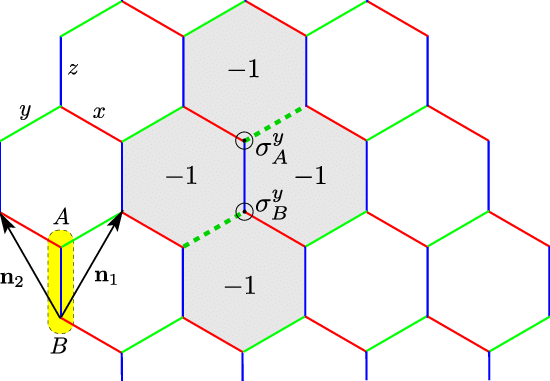}
  \caption{The KHM is formulated on the honeycomb lattice with a two-site cell (A and B).
The coupling pattern for the spin components is shown in red, blue, and green.
One can define conserved fluxes, which are $\pm 1$ throughout the lattice.
The picture is taken from Ref.~\cite{Knolle2014}}\label{fig_khm}
\end{figure}

{\it The Kitaev honeycomb model:}
In 2006, A. Kitaev introduced the KHM,
Ref.~\cite{Kitaev2006}. 
The microscopic model is one of spins, $s=1/2$, on the honeycomb lattice and reads
\begin{eqnarray}\label{eq_kitaevmodel}
  H=J_x \sum_{\color{red}{\langle i,j \rangle}}\sigma^x_{\color{red}i}\sigma^x_{\color{red}j}
  +J_y \sum_{\color{green}{\langle i,j \rangle}}\sigma^y_{\color{green}i}\sigma^y_{\color{green}j}
  +J_z \sum_{\color{blue}{\langle i,j \rangle}}\sigma^z_{\color{blue}i}\sigma^z_{\color{blue}j}\;.
  \label{eq_KHM}
\end{eqnarray}
Here, $\langle i,j \rangle$ denotes a sum over the nearest neighbors where bonds of different colors, red, green, or blue,
couple differently, while $\sigma^{x,y,z}$ are the standard Pauli matrices, see Fig.~\ref{fig_khm}.
It realizes an exact spin liquid ground state hosting exotic excitations, and in parts of its phase diagram, the elementary excitations are non-Abelian anyons.
The model can be solved exactly and features an effective theory corresponding to free Majorana fermions hopping on the honeycomb lattice.
Formally, this theory is equivalent to the tight-binding theory of graphene, albeit with Majorana degrees of freedom instead of actual fermions.

{\it Strain as a 'magnetic field':} For graphene, it has been shown that strain can mimic the effect of a magnetic field.
Specific patterns of strain lead to flat bands with a spacing reminiscent of Landau levels\cite{Guinea2010,Neek2013},
albeit without the topological properties due to the absence of time-reversal symmetry breaking. 
Following this strategy, one can emulate magnetic fields of gigantic strengths up to field equivalents of $250$ Tesla.   
Due to the formal equivalence between graphene and the zero flux sector of the Kitaev honeycomb model, the same also holds for the KHM~\cite{Rachel2016}.
The only caveat is that it is a priori not clear that the ground state is in the flux free sector since Lieb's theorem,
Ref.~\cite{Lieb1994}, only holds in systems with translation invariance.
However, this seems to be the case as demonstrated in~\cite{Rachel2016} and reinvestigated below. 
For our paper, the important feature of the 'Landau levels' is to quench the kinetic energy, and their loss of topology is irrelevant.
The flat bands lead to an effectively zero-dimensional problem as required for the SYK model.
 
{\it Experimental situation:} One of the reasons the KHM received a lot of interest was that in recent years the so-called iridates of first~\cite{oMalley2008,Abramchuk2017,Singh2012} and second generation ~\cite{oMalley2012,Kitagawa2018,Todorova2011,Roudebush2016,Takayama2015} could be identified as a material system in which the KHM could be realized, see \eg Ref.~\cite{Takagi2019} for a relatively recent overview.
At the moment, the most prominent candidate for hosting a Kitaev type spin liquid ground state may be the two-dimensional material $\rm{H}_3\rm{LiIr}_2\rm{O}_6$\cite{Kitagawa2018}.

Two types of perturbations that seem generically present in the candidate systems are a Heisenberg-type coupling and the $\gamma$-term~\cite{Rau2014,Katukuri2014,Yamaji2014},
which in the Majorana language assume the form of an interaction term not unlike the Coulomb interaction in graphene.
Usual studies of the KHM consider both types of perturbations detrimental to the quantum spin liquid physics,
although the properties are stable to small perturbations.
In this work, we will show how one may turn this nuisance into a desired feature that helps realize a version of the SYK model.

\subsection{The Kitaev honeycomb model and Majorana fermions}\label{sec:KHM}
Here, we review the KHM and its solution using a Majorana fermion representation of spin operators.
For a thorough discussion of the technical details, the reader is referred to Kitaev's original paper in Ref.~\cite{Kitaev2006} but also to later pedagogical reviews like Ref.~\cite{Mandal2020}.
It is worthwhile mentioning that the model is 'very forgiving' and can also be solved with an array of Jordan-Wigner type transformations~\cite{Feng2008,Chen2008,Kells2009} or parton constructions~\cite{Burnell2011}.

We first rewrite the KHM, Eq.~\eqref{eq_KHM}, in a more generic form as 
\begin{equation}
H_{K}=-\sum_{\left\langle i,j\right\rangle }J_{ij}\hat{\sigma}_{i}^{\alpha_{ij}}\hat{\sigma}_{j}^{\alpha_{ji}}\label{eq_H_spin_ham}
\end{equation}
where $\left\langle i,j\right\rangle $ is an ordered set of neighbors and $J_{ij}=J_{ji}$.
We assume that for each combination $i$,$j$, there is only one spin component, \ie, $\alpha_{ij}\neq\alpha_{il}$ if $j\neq l$.
The following solution in principle works on all trivalent lattices, meaning every lattice site is required to have at most three nearest neighbors which allows the use of any of three Pauli matrices at most once.
The standard setup of the KHM is shown in Fig.~\ref{fig_khm}, where the $\color{red}x$,
$\color{green}y$ and $\color{blue}z$ couplings are present on the three nonequivalent link-directions. 

The original solution by Kitaev~\cite{Kitaev2006} proceeded along the following lines:
in order to describe a local spin, he introduced two local fermions, implying the local Hilbert space dimension is four.
These two fermions are represented by four Majorana fermions $b^{x}$, $b^{y}$, $b^{z}$ and $c$.
The local Hilbert space splits into two sectors characterized by their fermion parity, even and odd. The dimension of each sector, respectively, is two.
One can define a parity operator $D=b^{z}b^{y}b^{x}c$ that distinguishes the two sectors and has eigenvalues $\pm1$.
In the extended local Hilbert space, one can represent the Pauli matrices in terms of the Majorana fermions according to
\begin{equation}
\hat{\sigma}_{i}^{\alpha}=\i\hat{b}_{i}^{\alpha}\hat{c}_{i}\;.\label{eq_KM_Defintion}
\end{equation}
While these matrices act in the four-dimensional local Hilbert space, they act like the standard Pauli matrices within the respective parity sectors and do not mix the two sectors.
It is straightforward to define a projection operator $P_{\pm}=\frac{1}{2}\left(1\pm D\right)$ projecting into the respective parity sectors. 

 Written in terms of Majorana fermions, Eq.~\eqref{eq_KM_Defintion} assumes the form
\[
H_{K} = \i\sum_{\left\langle i,j\right\rangle }J_{ij}\left(\i\hat{b}_{i}^{\alpha_{ij}}\hat{b}_{j}^{\alpha_{ji}}\right)\hat{c}_{i}\hat{c}_{j}\;.
\]
One proceeds to introduce the bond variables
$\hat{u}_{ij}=\i\hat{b}_{i}^{\alpha_{ij}}\hat{b}_{j}^{\alpha_{ij}}$ which square to one, \ie, $\hat{u}_{ij}^{2}=1$.
Consequently, $\hat{u}_{ij}$ has the eigenvalues $\pm1$.
All the $\hat{u}_{ij}$ commute among themselves as well as with any $\hat{c}_{k}\hat{c}_{l}$.
As a result, the $\hat{u}_{ij}$ commute with the Hamiltonian, meaning they are conserved and have independent eigenvalues.
In terms of these new variables, the Hamiltonian takes the form 
\begin{equation}
H_K =\i\sum_{\left\langle i,j\right\rangle }J_{ij}\hat{u}_{ij}\hat{c}_{i}\hat{c}_{j}\;,\label{eq_H_k_bonds}
\end{equation}
which is a bilinear in the operators $\hat{c}_{i}$ and $\hat{c}_{j}$.
In order to construct the full spectrum, in principle, one has to consider all configurations of the bond variables for the $\hat{u}_{ij}$ operators and calculate the remaining free hopping problem.
However, many of the different arrangements of the $\hat{u}_{ij}$ are gauge equivalent, and the only relevant quantity is the flux through a hexagon.
In a translationally invariant system, the ground state resides in the flux-free sector~\cite{Kitaev2006,Lieb1994} and the resulting model for the Majorana fermions is identical to the tight-binding problem in graphene. 

In this work, we concentrate on open systems. The boundary comes with two problems: the counting of degrees of freedom and, related, the fixing of a gauge.
As a technical aside, in Appendix~\ref{app:CountingDOFs}, we show that the counting of degrees of freedom works the same way as in the case with periodic boundary conditions in the bulk of the system.

\begin{figure}
\begin{centering}
\includegraphics[width=0.95\columnwidth]{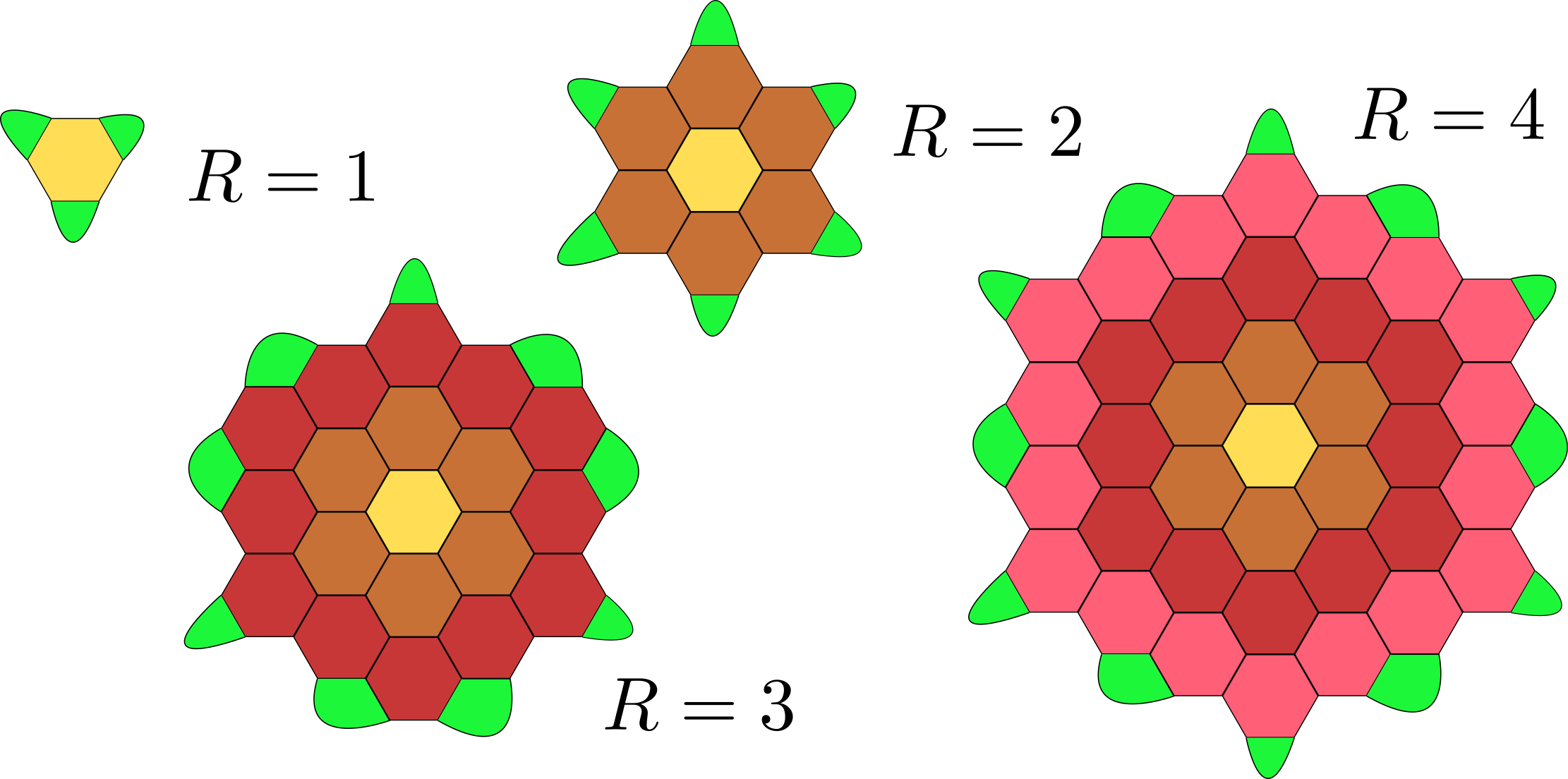}
\par\end{centering}
\caption{Sketch of a possible arrangement of ``fictitious'' plaquettes on the boundary of the Honeycomb flake for $R=1,2,3,4$ rings in the Honeycomb.
  There are $6R$ dangling bond Majoranas on the border which can be used to form  $3R$ plaquettes (green) on the border.
  Each bond plaque contributes an exact degeneracy - or strong zero-mode.
  \label{fig_BorderBonds}}
\end{figure}

The gauge equivalence breaks down at the edge of the system.
We find dangling bond variables that are not part of the Hamiltonian.
These dangling bonds can be used to form ``fictitious'' flux plaquettes that cost no energy,
thereby contributing to a massive degeneracy of states, not only in the ground state sector.
An illustration of how these bonds can be placed can be seen in Fig.~\ref{fig_BorderBonds}.
In this work, we ignore the existence of these states.
However, we note that their presence potentially has an effect on the perturbation theory sketched in Sec.~\ref{sec:Interaction}, and is left for future studies.

\subsection{The strained Kitaev model}\label{sec:Strain}

We consider a flake of the KHM subject to triaxial strain.
The flake consists of $R$ concentric rings, see Fig.~\ref{fig_BorderBonds}.

\begin{figure}
  \includegraphics[width=0.3\textwidth]{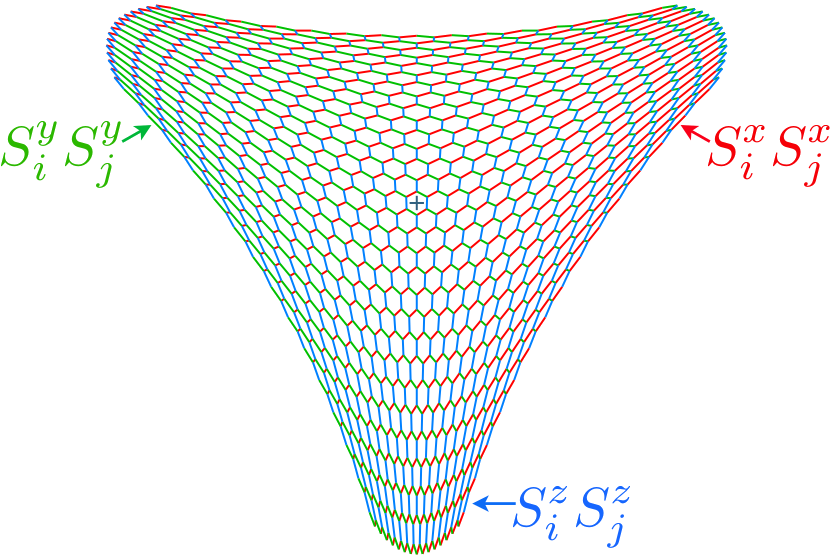}
  \caption{Trigonal strain applied to a flake of the KHM that leads to 'Landau levels'.
    Figure from Ref.~\cite{Rachel2016}.}\label{fig_strain}
\end{figure}

\begin{figure}
  \begin{centering}
    \includegraphics[width=0.95\columnwidth]{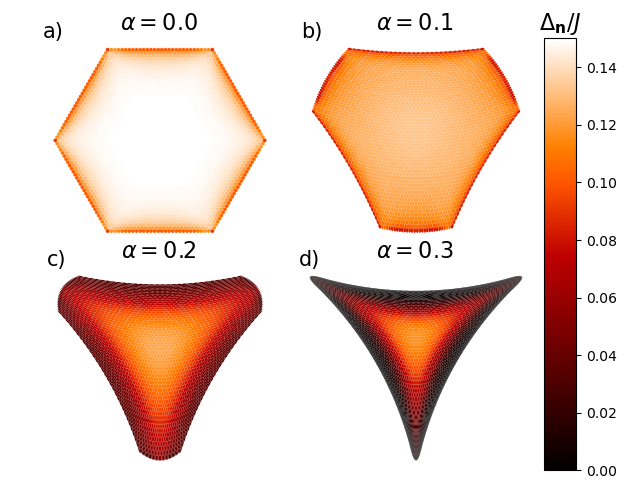}
    \end{centering}
  \caption{Layout of the 1-flux energy landscape, $\Delta_{\bf n}$, for $R=30$ rings.
  For stronger strain, $\alpha=0.3$, the flux gap closes on the boundary of the strained Honeycomb flake, while  for weaker strain $\alpha\leq0.2$ it remains open.}
  \label{fig_One-Flux-landscape}
\end{figure}

Before applying strain, we assume spatially uniform couplings, \ie, $J_{ij}=J$.
Applying a strain pattern as shown in Fig.~\ref{fig_strain} leads to 'Landau levels' of Majorana fermions~\cite{Rachel2016}, albeit without the topological properties of Landau bands.
We closely follow the approach in Ref.~\cite{Rachel2016} and introduce triaxial strain (also see Refs.~\cite{Guinea2010,Neek2013})
such that each lattice point $\vec{r}_{i}=\vec{R}_{i}$ gets displaced to $\vec{r}_{j}\to\vec{R}_{i}+\vec{U}_{i}$, where 
\begin{equation}
\vec{U}_{i}=\vec{U}\left(x_{i},y_{i}\right)=\bar{C}\left(2x_{i}y_{i},x_{i}^{2}-y_{i}^{2}\right),\label{eq_Displacement}
\end{equation}
is the space dependent displacement.
We parameterize the 'strength' of strain according to
\[
\bar{C}=\frac{\alpha}{a_0R}
\]
where $\alpha$ is an intensive quantity that, independent of the system size $R$,
characterizes the shape of the flake in the sense that different sizes can be scaled onto each other if they possess the same $\alpha$. 
To illustrate this let us move a point on the $x$-axis from the position $\vec{r}=\left(nRa_0,0\right)$
to some other point $\vec{r}^\prime=\left(nRa_0,mRa_{0}\right)$.
We keep $n$ and $m$ fixed (with $m$ depending indirectly on $n$) and independent of the system size $R$.
For this move we need a displacement $\vec{U}=\left(0,mRa_{0}\right)$.
Comparing to equation \eqref{eq_Displacement} where we set $y=0$
and $x=nRa_0$ we get $mRa_0=\bar{C}\left(nRa_0\right)^2$
such that $\bar{C}=\frac{m}{n^{2}Ra_{0}}$ and we can identify $\alpha=\frac{m}{n^{2}}$.
For a visualization of the strain-shape dependence on $\alpha$ please look ahead to Fig.~\ref{fig_One-Flux-landscape} where this is shown explicitly.

\begin{figure}
  \begin{centering}
    \includegraphics[width=0.95\columnwidth]{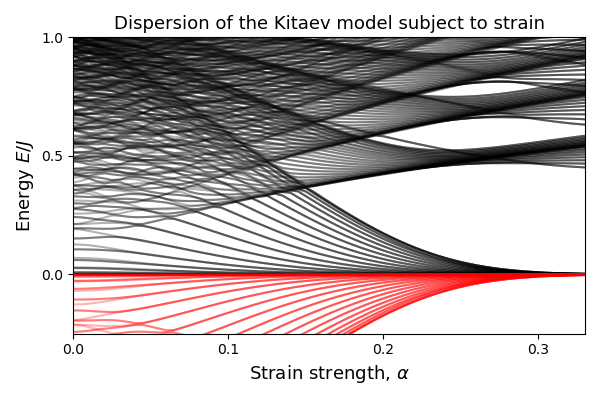}\\
    \par\end{centering}
    \caption{Energy levels for a system with $R=20$ rings as a function of $\alpha$.
      At $\alpha\approx0.15$ a gap is opening and at $\alpha=0.3$ the lowest band is flat.
     The $\alpha$-dependence of the qualitative features in the plot is independent of the system size $R$.}
    \label{fig_Landau_bands}
\end{figure}

Under the applied strain the coupling constants get modified to lowest order according to 
\begin{equation}
J_{ij}=J\left[1-\beta\left(\left|\vec{\delta}_{ij}\right|\frac{1}{a_{0}}-1\right)\right]\label{eq_J_ij}
\end{equation}
 where $\vec{\delta}_{ij}=\vec{r}_{i}-\vec{r}_{j}=\left(\vec{R}_{i}+\vec{U}_{i}\right)-\left(\vec{R}_{j}+\vec{U}_{j}\right)$
 is the relative distance after displacement.
 Throughout this paper we choose $\beta=1$, but it is in principle system specific.
 
Due to the bipartite structure of the honeycomb lattice we can split the Majorana fermions into two groups, $A$ and $B$, and rewrite Eq.~\eqref{eq_H_k_bonds} as 
\begin{equation}
H_{K} =i \sum_{i\in A,j\in B}\hat{c}_{i}^{A}C_{ij}\hat{c}_{j}^{B}\;,\label{eq_H_k_bipartite}
\end{equation}
where $C_{ij}=J_{ij}\hat{u}_{ij}$ is the kernel of the free Hamiltonian.
Since $A$ and $B$ are distinct sets of Majoranas, there is no double-counting and no factor of $\frac{1}{2}$.
This Hamiltonian is readily solved by means of a Single-Value-Decomposition (SVD) where $C_{ij}=\sum_{\lambda}U_{i\lambda}S_{\lambda}V_{\lambda j}^{T}$.
Inserting the SVD, we directly obtain
\begin{equation}
H_K = \sum_{\lambda}S_{\lambda}\left(\i a_{\lambda}^{A}a_{\lambda}^{B}\right)\label{eq_H_k_SVD}
\end{equation}
where $S_{\lambda}\geq 0$ and
\begin{align}
a_{\lambda}^{A} & =\sum_{i\in A}\hat{c}_{i}^{A}U_{i\lambda}\;,\nonumber \\
a_{\lambda}^{B} & =\sum_{j\in B}V_{\lambda j}^{T}\hat{c}_{j}^{B}\;,\label{eq_a_of_c}
\end{align}
are the Majorana degrees of freedom.
The ground state is given by the physical state $\ket 0$ defined through $a_{\lambda}^{B}\ket 0=\i a_{\lambda}^{A}\ket 0$
(or equivalently $\i a_{\lambda}^{A}a_{\lambda}^{B}\ket 0=-\ket 0$) for all $\lambda$.
The ground state energy is given by $\epsilon_{\text{GS}}=-\sum_{\lambda}S_{\lambda}$.
We note that, due to the constraints coming from the projective construction, we have either 'only even' or 'only odd' number of Majorana fermions in the physical state.
The parity, however, depends on both the configuration of fluxes and on the specific couplings in $C_{ij}$.
Consequently, the parity is directly related to the signs of the determinants of $U$ and $V$.

Under strain, the Kitaev model hosts flat bands~\cite{Rachel2016},
just like graphene, see Ref.~\cite{Guinea2010, Neek2013}.
This is illustrated in the upper panel of Fig.~\ref{fig_Landau_bands} where flat bands develop as a function of $\alpha$ (note that this is the spectrum in the zero flux sector).
Around $\alpha=0.15$ a gap in the spectrum opens up, and at $\alpha=0.3$, the lowest band flattens completely to form a Landau level like band.
At about the same $\alpha$, excited states of the same Landau level type are separated by the analogue of the cyclotron frequency.

\subsection{The fate of the flux gap}\label{sec:Fluxgap}
For the translationally invariant KHM, the ground state can be shown to be in the flux-free sector following Lieb's theorem~\cite{Lieb1994}.
A finite flake under strain does not possess translational invariance, which naturally begs the question of whether the ground state is still in the flux-free sector.
For this purpose we look at the clean system with $n=0,1,2,3,\dots$
fluxes and then compute the ground state energy within each flux configuration ${\bf n}$.
For each flux configuration ${\bf n}$ we record the ground state energy $\epsilon_{\text{GS},\bf n}=-\sum_{\lambda}S_{\lambda}$,
and the corresponding flux gaps $\Delta_{\bf n}=\epsilon_{\text{GS},\bf n} - \epsilon_{\text{GS},0}$.\\

{\it The 1-flux gap}:
To begin, we map out the flux gap to all configurations of the entire 1-flux sector, for various $\alpha$, see Fig.~\ref{fig_One-Flux-landscape}.
The images are to be read as follows: the darker the color, the smaller the flux gap if the respective flux is located in said position. 
We find that with fluxes in the center of the flake, not unexpectedly, the flux-gap is almost unchanged compared to the infinite system,
whereas for fluxes at the stretched boundary, the flux-gaps significantly reduce and for $\alpha=0.3$ even seem to disappear.
We speculate that the main reason that the boundary has most of the gap closing configurations,
is that it is more strongly affected by the advent of strain,
\ie, distances between sites get deformed more, and there are generically more low-energy configurations.

\begin{figure}
  \begin{centering}
    \includegraphics[width=0.95\columnwidth]{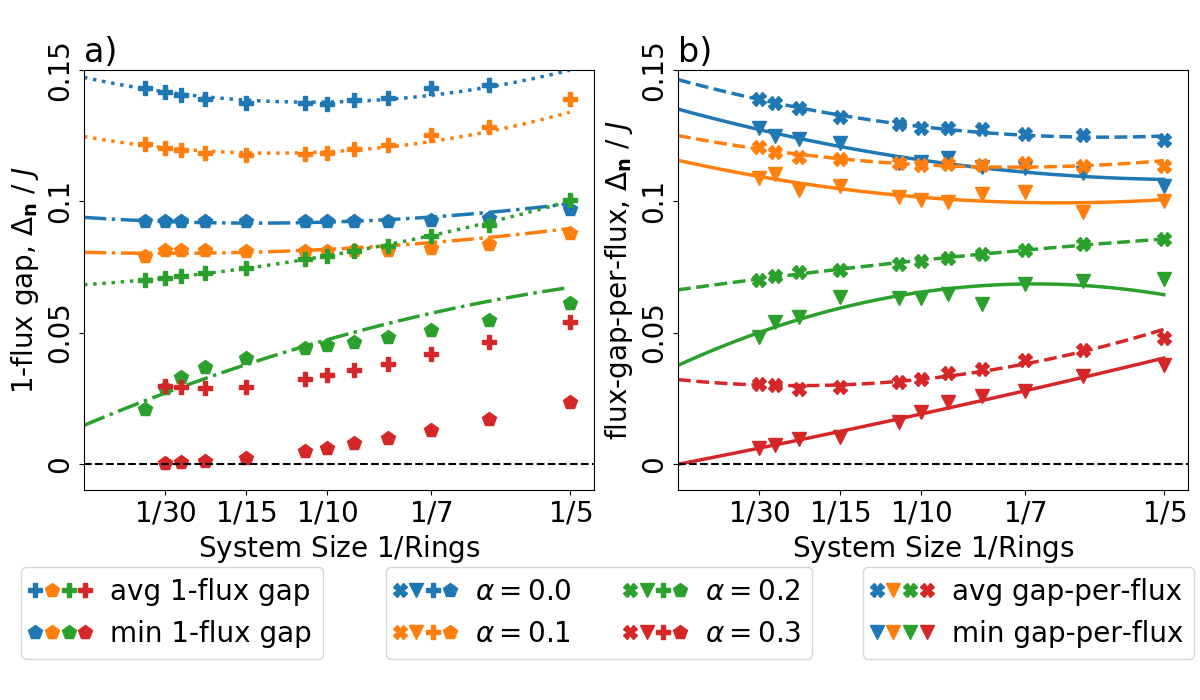}
  \end{centering}
  \caption{a) Scaling of the minimal and average 1-flux gap $\Delta_{\bf n}$ again $1/R$ for a range of strain shapes $\alpha$.
    b) Scaling of the minimal and average flux-gap-per-flux $\Delta_{\bf n}/n$ again $1/R$ for the same $\alpha$.
    For strong strain $\alpha=0.3$, the gap closes in the thermodynamic limit $R\to\infty$, but not for weaker strain $\alpha\leq0.2$.
      }\label{fig_GSScaling}
\end{figure}

To quantify these observations, we perform finite-size scaling of the minimal and average 1-flux-gaps, see Fig.~\ref{fig_GSScaling}a).
In the panel, we measure the minimal and average gap from the zero-flux sector to the one-flux sector as a function of $1/R$.
We scan over all possible flux configurations:
We record both the smallest and the average flux gap, $\Delta_{\bf n}$.
When appropriate, we apply a quadratic fit to take into account that the number of sites grows quadratically with $R$, to determine the thermodynamic scaling.

Let us first look at the case of strong strain, $\alpha=0.3$ ({\color{red} red}).
There we find that the 1-flux gap approaches zero faster than quadratically as a function of $1/R$.
This indicates that it will, in practice, close before the thermodynamic limit is reached.
The available data does {\it not} suggest that the 1-flux sector would contain the global ground state, though.
However, the same is not true for the average gap to the 1-flux sector,
where there seems to be a spread of $\Delta_E\approx0.03 J$ in the thermodynamic limit.

For less strain $\alpha<0.3$, the 1-flux-gap seems to remain open in the thermodynamic limit.
We note, however, that $\alpha=0.2$ ({\color{Green} green}) has a downward trend and may close if larger system sizes were to be added to the analysis.

Note that since the average 1-flux-gap does not go to zero, the 1-flux sector will always have some gaped configurations for all investigated values of $\alpha$.

Comparing the flux gap to the band width in Fig.~\ref{fig_Landau_bands}, we see that there seems to be a sweet spot around $\alpha\sim0.2$ where the flux gap between the \textbf{zero-flux } bands and the \textbf{1-flux} sector is larger than the \textbf{zero-flux} bandwidth.

Finally, since the gap closing appears at the boundary, it remains an open question what the fate of the flux-gap is in the presence of disorder on the boundary.
An in-depth study of this case is left to the future.
Here, we assume that the effect will be subleading concerning bulk properties.

\begin{figure}
  \begin{centering}
    \includegraphics[width=0.95\columnwidth]{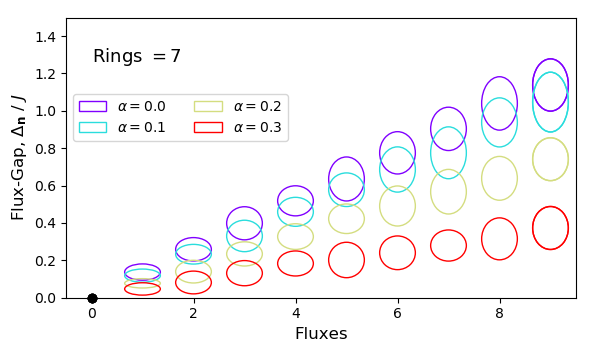}
    \par\end{centering}
    \caption{Distribution of flux gaps in sectors with $n$ fluxes compared with the 0-flux sector, on size $R=7$ and increasing strain.
      A small flake is shown here to allow for a large number of fluxes.
      The top and bottom of each ellipse mark the smallest and largest flux gap at a fixed number of fluxes.
      The flux-gap-per-flux is almost constant and is observed to decrease as the strain increases, just like the 1-flux gap.
      See Fig.~\ref{fig_GSScaling}.
    }\label{fig_GS_Scan}
\end{figure}

{\it Flux-gap-per-flux}:
Having studied the behavior or the 1-flux gap, we now turn our attention to configurations with several fluxes.
For $n=1$ we could scan over all the flux-configurations but for $n>1$ the numerical cost explodes,
as there are too many different configurations with a fixed number of fluxes.
Thus in what follows for $n\geq2$ we will randomly select a large number of flux-configurations for each $n$.

We begin by checking whether a ``new'' ground state could be found in sectors with multiple fluxes.
In Fig.~\ref{fig_GS_Scan}, for illustrational purposes, we show how the flux-gap distribution depends on the number of fluxes, $n$,
for a small system with $R=7$ rings and increasing strain, $\alpha$.
We see that as $\alpha$ increases, the global ground state is still always in the \textbf{zero flux} sector.
Furthermore, each extra flux ads on average the same energy $\Delta E$.
For $\alpha=0.0$ it is $\Delta E\sim0.1$ while it is $\Delta E\sim0.05$ when $\alpha=0.3$.
Thus the flux-gap-per-flux decreases with increasing strain, just like the 1-flux gap.

In Fig.~\ref{fig_GSScaling}b) we perform a scaling analysis also for the minimal and average flux-gap-per-flux.
Here the number of fluxes that we use depends on the system size. 
In this approach, we scan over several different flux-sectors and for each flux sector, sample a random selection of flux configurations, 
again since the total space of flux-configurations becomes too large otherwise.
We then perform a linear fit through the average and minimal flux-gap and plot it against the inverted system size.

For strong strain, $\alpha=0.3$ ({\color{red} red}), we find that the flux-gap-per-flux also goes to zero in the thermodynamic limit, but this time linearly in $1/R$.
The average gap, however, remains open at $0.02 J$ just as in Fig.~\ref{fig_GSScaling}a).
Note that for less strain, $\alpha\leq0.2$, the minimal flux-gap-per-flux seems to remain finite in the thermodynamic limit.

From this analysis, it appears that the global ground state remains in the zero-flux sector for small systems and not too strong strain,
although the flux-gap appears to close if the strain becomes too strong.
Thus, for a mesoscopic flake, there is a trade-off to be made between the amount of strain and the system size,
with smaller system sizes being able to sustain more strain while still keeping the flux-gap open.

\section{The Effective Interacting Hamiltonian}\label{sec:Interaction}

The purpose of this section is to investigate the role of perturbations in the ideal KHM.
For the following discussion, it is important that there is a finite flux gap,
which we argued above to be present for finite size systems with not too strong strain. 
We consider the projective physics of all perturbations in the zero flux sector.
A more in-depth discussion of this aspect of our work is planned for the future but beyond the scope of the present paper.

There are two types of generic perturbations in actual realizations of the KHM. One is of the Heisenberg type, meaning 
\begin{eqnarray}
V_H=J_H \sum_{\left \langle i,j  \right\rangle}  \sum_{\mu=x,y,z}\hat{\sigma}^\mu_i \hat{\sigma}^\mu_j \;.
\end{eqnarray}
Furthermore, there is a cross-term called the $\gamma$-term, which reads
\begin{eqnarray}
V_{C}=J_C\sum_{\left\langle i,j\right\rangle}\sum_{\nu,\gamma\in x,y,z}\left|\epsilon^{\mu_{i,j}\nu\gamma}\right|\sigma_{i}^{\nu}\sigma_{j}^{\gamma}\;,
\end{eqnarray}
where $\epsilon^{\mu\nu\gamma}$ is the Levi-Civita antisymmetric tensor such that $\nu,\gamma$ are the complementary indexes to the Kitaev spin component $\mu_{i,j}$.

\begin{figure}[t]
\begin{centering}
\includegraphics[width=0.4\paperwidth]{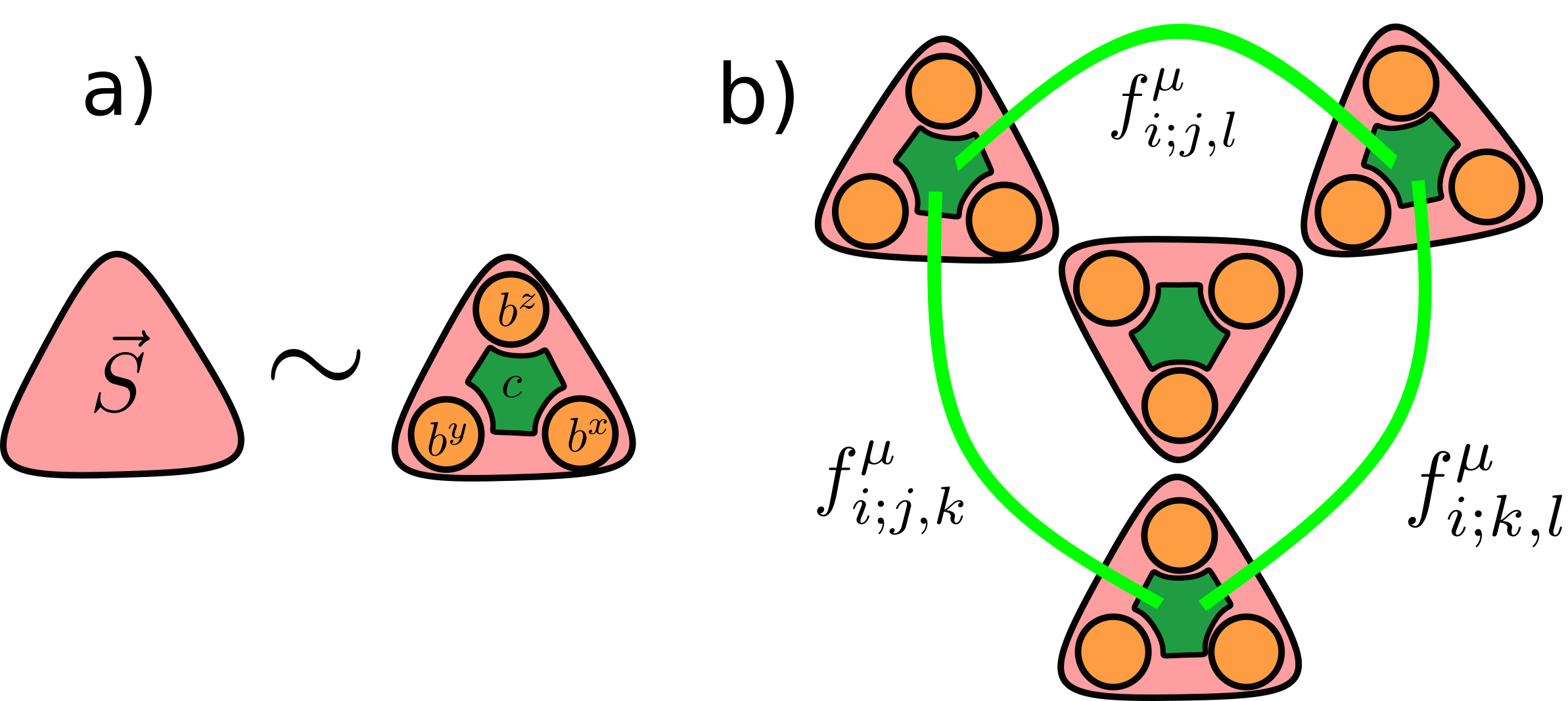}
\par\end{centering}
\caption{a) The standard Kitaev construction of mapping a spin degree of freedom to four majoranas $\sigma^\mu = \i b^\mu c$.
  b) Structure of the first correction term to $\sigma^\mu$ given by eqn.~\ref{eq_Sigma}.
  Note that the pair of majoranas $c_jc_k$ are on the same sub-lattice.
  \label{fig_Interaction_Combintations}}
\end{figure}

We now follow the symmetry-based analysis of Ref.~\cite{Song2016} where it was shown that at low energy,
after a projective analysis of the perturbations, the spins effectively develop components that are bilinear in the $c$-Majorana fermions and take the form
\begin{equation}
\tilde{\sigma}_{i}^{\mu}=\i Z_{i}c_{i}b_{i}^{\mu}+\frac{\i}{2}\sum_{j,k\in N_{i}}f_{i;j,k}^{\mu}c_{j}c_{k}+\cdots.\label{eq_Sigma}
\end{equation}
The coupling constants are asymmetric, $f_{i;j,k}^{\mu}=-f_{i;k,j}^{\mu}$,
$f_{i;j,j}^{\mu}=0$ and couple Majorana fermions on sites adjacent to the physical spin, see Fig.~\ref{fig_Interaction_Combintations}.
Here $N_{i}$ is the set of neighbors of site $i$, and $j,k$ are thus the neighbors of $i$. 
Note that $\left|Z_{i}\right|<1$ is a renormalization constant that ensures $\left(\tilde{\sigma}_{i}^{\mu}\right)^{2}=1$.
In the presence of both Heisenberg and $\gamma$-terms, this is the lowest nontrivial perturbation that arises on the Majorana level.

Reinserting Eq.~\eqref{eq_Sigma} for $\tilde{\sigma}_{i}^{\mu}$ into Eq.~\eqref{eq_H_spin_ham}, we obtain two terms (the term linear in $b$ can be discarded in the low energy limit)
\begin{equation}
H= \sum_{i\in A}\sum_{j\in N_{i}}J_{ij}Z_{i}Z_{j}b_{i}^{\alpha_{ij}}b_{j}^{\alpha_{ij}}c_{i}^{A}c_{j}^{B} + \HamI\;.
\end{equation}
This is just a renormalized version of the standard Kitaev model,
albeit dressed, whereas the second term
\begin{widetext}
\begin{equation}
\HamI=-\frac{1}{4}\sum_{i\in A}\sum_{j\in N_{i}}\sum_{l_{i},k_{i}\in N_{i}}\sum_{l_{j},k_{j}\in N_{j}}J_{ij}f_{i;l_{i},k_{i}}^{\alpha_{ij}}f_{j;l_{j},k_{j}}^{\alpha_{ij}}c_{l_{i}}^{B}c_{k_{i}}^{B}c_{l_{j}}^{A}c_{k_{j}}^{A},\label{eq_H_I}
\end{equation}
describes the effective interactions between the Majorana fermions and is the term we concentrate on.
\end{widetext}

For simplicity, we assume $Z_{i}=Z$ to be constant such that the rescaling is uniform within the low energy sector.
Similarly, we assume that the couplings $J_{ij}$ follow Eq.~\eqref{eq_J_ij}.
The free Majorana spectrum is identical (albeit rescaled) to that of the pure Kitaev model
meaning we can write the Hamiltonian as $H=Z^2 H_{K}+\HamI$ where $H_{K}$ is given by \eqref{eq_H_k_SVD}.
In this expression, the Majorana fermions are again related to the 'original' ones according to \eqref{eq_a_of_c}, and $\HamI$ is the interaction Hamiltonian introduced in Eq.~\eqref{eq_H_I}.
In terms of the original spins, the function $f_{ijk}^{\mu}$ picks out pairs of neighbors around the site $i$.
Since the Kitaev model only has nearest-neighbor spin interactions, this means that the interaction Hamiltonian, Eq.~\eqref{eq_H_I},
has nine configurations that contribute to every $i$,$j$ pair.

To represent Eq.~\eqref{eq_H_I} in terms of the $a$-Majoranas that diagonalize $\mathcal{H}_{0}$ we invert the relations found in Eq.~\eqref{eq_a_of_c} according to 
\begin{align}
c_{i}^{A} & =\sum_{n=1}^{N_{A}}a_{n}^{A}U_{ni}^{T}\nonumber \\
c_{j}^{B} & =\sum_{m=1}^{N_{B}}V_{jm}a_{m}^{B}\label{eq_c_of_a}\;.
\end{align}
 Note that in general $N_{A}\neq N_{B}$, where $N_A$ ($N_B$) is the number of $A$-sites ($B$-sites) in the system. However,
as long as we work with $U$ and $V^{T}$ from the full SVD, then both $U$ and $V$ can always be inverted. More details can be found in Appendix~\ref{app:Majoran-svd}.
 
 We proceed to insert Eq.~\eqref{eq_c_of_a} into Eq.~\eqref{eq_H_I} to get
\begin{align*}
  \HamI = \frac{1}{4}\sum_{n_1,n_2=1}^{N_A}\sum_{n_3,n_4=1}^{N_B}
  J_{n_1,n_2;m_3,m_4}a_{m_3}^Ba_{m_4}^Ba_{n_1}^Aa_{n_2}^A\;,
\end{align*}
where
\begin{align*}
  J_{n_{1},n_{2};m_{3},m_{4}} = & \sum_{i\in A}\sum_{j\in N_{i}}\sum_{l_{i},k_{i}\in N_{i}}\sum_{l_{j},k_{j}\in N_{j}}\\
  & J_{ij}f_{i;l_{i},k_{i}}^{\alpha_{ij}}f_{j;l_{j},k_{j}}^{\alpha_{ij}}V_{l_{i}m_{3}}V_{k_{i}m_{4}}U_{n_{1}l_{j}}^{T}U_{n_{2}k_{j}}^{T}.
\end{align*}

The coupling constant $J_{n_1,n_2;m_3,m_4}$ can be organized as a sum over local terms by writing it as 
\begin{align}
J_{n_1,n_2;m_3,m_4} = \sum_{i\in A}\sum_{j\in N_{i}}J_{ij}\Pi_{m_{4},m_{3}}^{\left(i;j\right)}\Gamma_{n_{1},n_{2}}^{\left(j;i\right)}\,,\label{eq_j_of_pi}
\end{align}
 where 
\begin{align}
\Pi_{m_{4},m_{3}}^{\left(i;j\right)} & =\sum_{l_{i},k_{i}\in N_{i}}f_{i;l_{i},k_{i}}^{\alpha_{ij}}V_{l_{i}m_{3}}V_{k_{i}m_{4}}\;,\label{eq_PI_and_Gamma}\\
\Gamma_{n_{1},n_{2}}^{\left(j;i\right)} & =\sum_{l_{j},k_{j}\in N_{j}}f_{j;l_{j},k_{j}}^{\alpha_{ij}}U_{n_{1}l_{j}}^{T}U_{n_{2}k_{j}}^{T}\;.
\end{align}
 We note that these terms explicitly enforce the anti-symmetry
\[
J_{n_{1},n_{2};m_{3},m_{4}}=-J_{n_{2},n_{1};m_{3},m_{4}}=-J_{n_{1},n_{2};m_{4},m_{3}},
\]
 since $\Gamma_{n_{1},n_{2}}^{\left(i;j\right)}=-\Gamma_{n_{2},n_{1}}^{\left(i;j\right)}$
and $\Pi_{m_{4},m_{3}}^{\left(j;i\right)}=-\Pi_{m_{3},m_{4}}^{\left(j;i\right)}$.
Note that, since $f_{i;jk}^{\alpha_{ij}}$ is anti-symmetric in $j,k$, it follows $\Pi_{m,m}^{\left(i;j\right)}=\Gamma_{n,n}^{\left(j;i\right)}=0$.
Thus we can further simplify $\HamI$ to 
\begin{equation}
\HamI=\sum_{n_{1}<n_{2}}^{N_{A}}\sum_{m_{3}<m_{4}}^{N_{B}}J_{n_{1},n_{2};m_{3},m_{4}}a_{m_{3}}^{B}a_{m_{4}}^{B}a_{n_{1}}^{A}a_{n_{2}}^{A}\;.\label{eq_H_i_short}
\end{equation}

We find that while this Hamiltonian looks reminiscent of the SYK model,
it inherits the bipartite nature of the original KHM. Therefore,
we refer to it as b-SYK (bipartite). This Hamiltonian was introduced in Eq.~\eqref{eq_bSYK} for random couplings.
The properties of this model for random couplings are related but different from the SYK model and are explored in a parallel paper~\cite{Fremling2021}.
In the following, we study the structure of the couplings.

\begin{figure*}[t]
  \begin{centering}
    \includegraphics[width=0.95\linewidth]{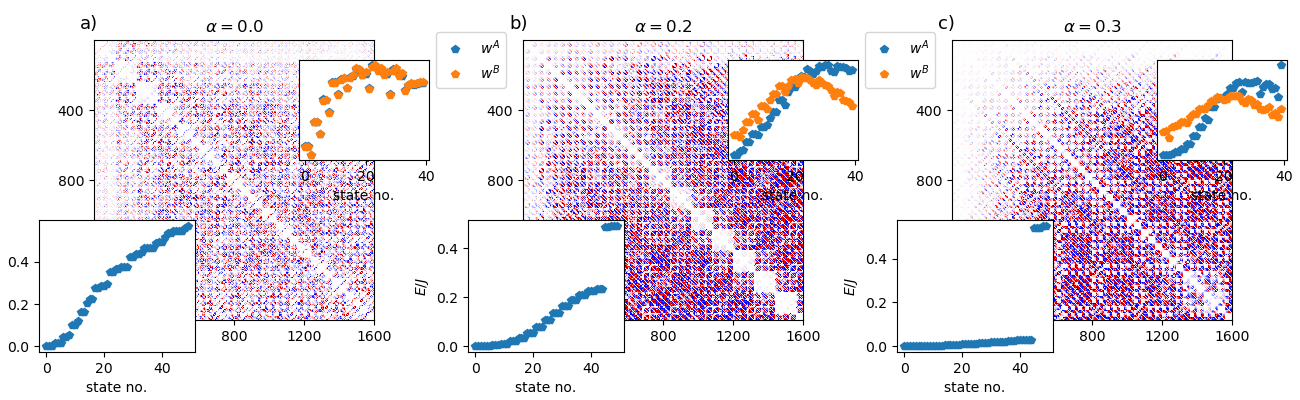}
  \end{centering}
  \caption{
    \emph{Main figure}: Visualization of the elements of $J_{n_1,n_2;m_3,m_4}$ by comcaptifying the indexes as as $L = (n_1,m_3)$ and $R = (n_2,m_4)$ to form the matrix $J_{L,R}$.
    The three panels show for $R=15$ rings a) an unstrained system,
b) an $\alpha=0.2$-strained system and c) an $\alpha=0.3$-strained system.\\
    \emph{Lower left inset}: The single particle majorana energies.
For the strained systems in b-c), there is a clear energy gap that is not present in the unstrained system a).\\
    \emph{Upper right inset}: Total interaction weight, $w$ for the two species of Majoranas defined in eqn.~\ref{eq_to_int_weight}.
    Note how the lowest energy Majoranas tend to have smaller interaction coefficients than the higher energy Majoranas.
    In the unstrained system, the A-B sublattice symmetry is not broken; therefore, $w^A = w^B$ there.
    \label{fig_Hop-Elem_flatten}}
\end{figure*}

\begin{figure*}[tb]
  \begin{centering}
    \includegraphics[width=0.95\linewidth]{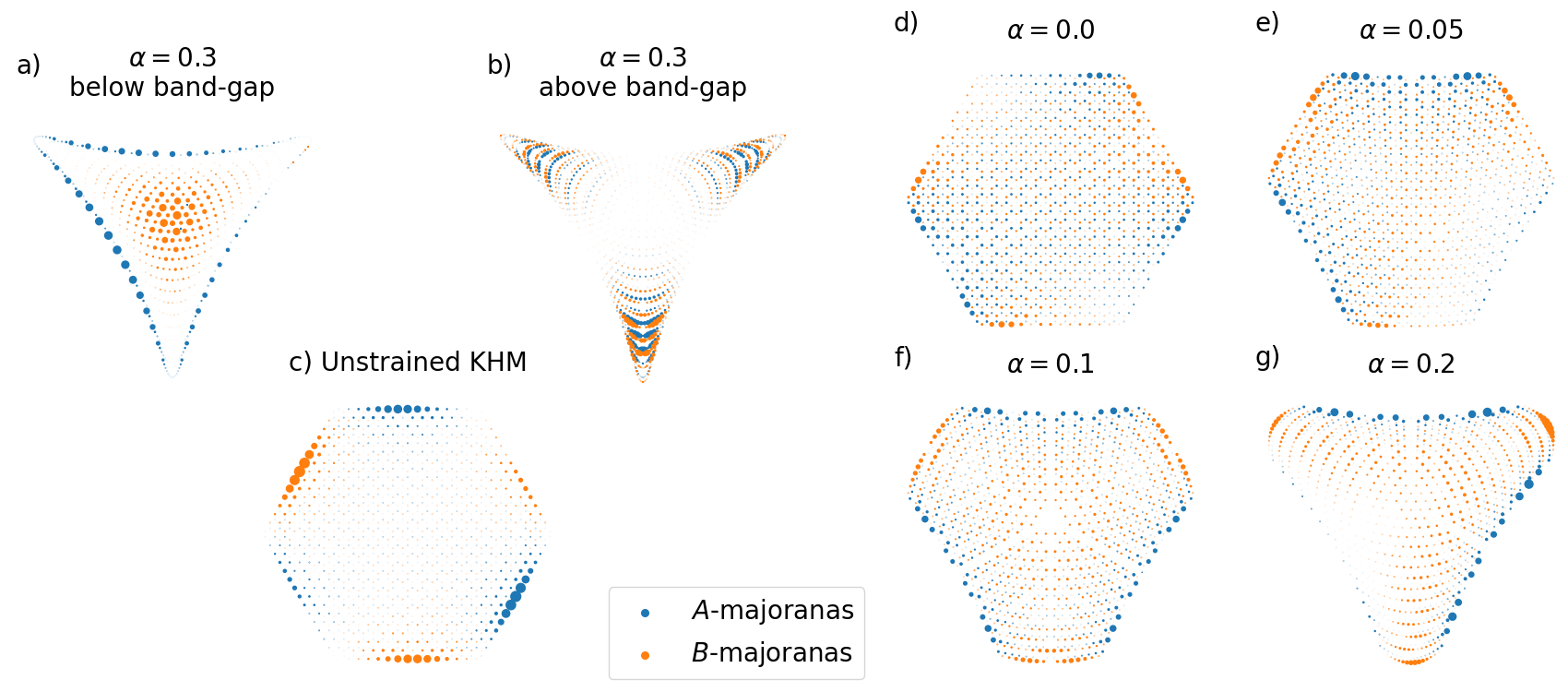}
    \par\end{centering}
    \caption{
      Illustrations of the distribution of $A$- and $B$-Majoranas for various amounts of strain. Here a system of $R=15$ rings is used.
    The area of the circles is proportional to density, ``$|\psi|^2$''.
      For a) and c), the lowest energy state is shown; for d-g), the 10:th lowest energy state is depicted.
      Note how the application of strain causes the $A$ and $B$ orbitals to separate below the bandgap, whereas the orbitals are mixed above the bandgap. 
      \label{fig_StateStructure}}
\end{figure*}

\section{Numerical Characterization of $\HamI$}\label{sec:Elems}

\subsection{The interaction elements $J_{n_1,n_2;m_3,m_4}$ }
In this section, we attempt to derive the interaction Hamiltonian in Eq.~\eqref{eq_H_i_short} from microscopics.
Unfortunately, calculating the structure factors $f_{i;j,k}^{\omega}$ requires perturbation theory to a very high order which at the moment is beyond the scope of this paper.
In order to model the interactions and the disorder,
we make the highly simplifying assumption that (i) $f_{i;j,k}^{\omega}=\pm f$ is a constant.
(ii) We assign the sign of $f$ randomly for each $f_{i;j,k}^{\omega}$,
since we have no a priory knowledge of the sign structure for $f$. 
While we impose the structure factors with a random sign, we consider the actual wave-functions of the strained flake according to Eq.~\eqref{eq_a_of_c} for the calculation of the interaction elements.
$J$ is anti-symmetric in both $m$ and $n$ indexes; thus, the interaction Hamiltonian does not hide any 'quadratic' terms in the $c$-Majorana fermions.
As a consequence, we do not need to consider the potential backreaction on the band structure from $\HamI$.

\subsubsection{Visualization and interaction weight, $w$}
To visualize the elements of the rank four tensor that is $J_{n_1,n_2;m_3,m_4}$ we compactify $L = (n_1,m_3)$ and $R = (n_2,m_4)$ to form the matrix $J_{L,R}$ wich can be seen in Fig.~\ref{fig_Hop-Elem_flatten} for the unstrained system (left) as well as a strained system with $\alpha=0.2$ (middle) and $\alpha=0.3$ (right).
In Fig.~\ref{fig_Hop-Elem_flatten}, $J$ has been sampled with the dimensions $(N\times M)\times (N\times M)$ such that $L = n_1M + m_3$ and $R=n_2M + m_4$.
In a disordered system, $\left|M-N\right|=|N_A-N_B|$ is equal to the site imbalance between the $A$ and $B$ sub-lattices,
and also equal to the number of exact zero modes related to dangling Majorana fermions.

In Fig.~\ref{fig_Hop-Elem_flatten}a) we observe that for an unstrained system there is almost no correlation between the different elements,
\ie, there is very little visible structure, except for the very lowest states (in the upper left corner).
To quantify this, we compute the \emph{total interaction weight} $w$ for each Majorana mode $a^A$ and $a^B$ that is sampled.
This measure simply sums up the total weight of all the interaction elements as
\begin{align}
  w^A_n =& \sum_{n_2,m_3,m_4}^{N_{\mathrm sampled}} \left|J_{n,n_2;m_3,m_4}\right| \\
  w^B_m =& \sum_{n_1,n_2,m_3}^{N_{\mathrm sampled}} \left|J_{n_1,n_2;m_3,m}\right|\;,\label{eq_to_int_weight}
\end{align}
and is shown in the inset of Fig.~\ref{fig_Hop-Elem_flatten}.
For the unstrained system we find that $w^A_n=w^B_n$ and that the first $n\sim 10$ modes have a significantly lower interaction weight than the other modes.

For the strained system (Fig.~\ref{fig_Hop-Elem_flatten}b), one can see that the interaction elements tend to get stronger as one moves from the upper left to the lower right.
This can be interpreted as states at lower energies tend to scatter significantly less than states at higher energy.
Again, this is corroborated by the $w$-measure, where the low energy states have a noticeably lower weight than the higher energy ones.
We believe that the main reason for this is that the $A$-Majorana modes live predominantly in the bulk, whereas the $B$-Majorana modes live close to the boundary for low energy states.
For an illustration of this, see Fig.~\ref{fig_StateStructure}.

It is also clearly visible that $w^A_n\neq w^B_n$, which is a direct consequence of the sub-lattice symmetry being broken by the applied strain.

\begin{figure}[t]
  \begin{centering}
    \includegraphics[width=0.95\linewidth]{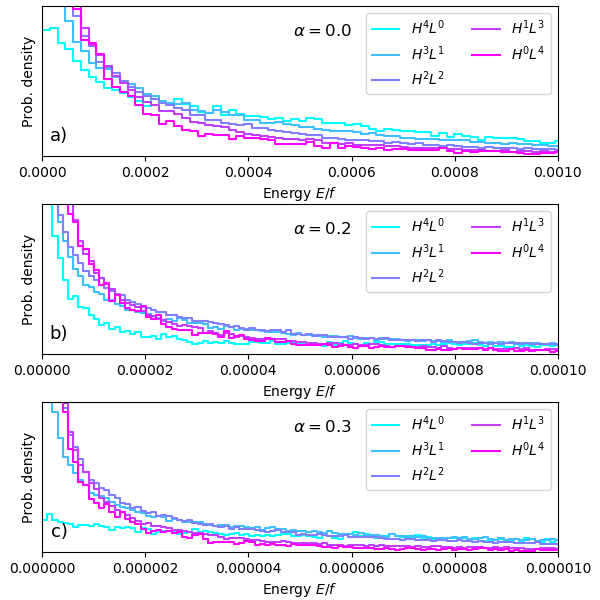}
    \par\end{centering}
    \caption{Distribution of interaction elements $J_{n_1,n_2;m_3,m_4}$ with and without strain for a system of $R=15$ rings, for certain partitions of $J$.
      The systems are the same as those depicted in Fig.~\ref{fig_Hop-Elem_flatten}.
      The $N_{\mathrm sampled}=40$ states are split into 20 ``high energy'' majoranas and 20 ``low energy'' majoranas.
      The label $H^nL^m$ should then be read as the amplitudes of all scattering events between $n$ ``high energy'' Majoranas and $m$ ``low energy'' Majoranas.
      In all three panels, one can see that $H^4L^0$ has the largest fraction of large interaction elements, whereas $H^0L^4$ has the lowest fraction.
    \label{fig_ElemStats}}
\end{figure}

\subsubsection{The distribution of $J_{n_1,n_2;m_3,m}$}
Let us briefly look at the distribution of the interaction elements.
In Fig.~\ref{fig_ElemStats}, the distribution of the interaction elements is plotted for various partitions of the data.
In this figure we split the $N_{\mathrm sampled}=40$ states into 20 ``high energy'' states and 20 ``low energy states''.
The label $H^nL^m$ should then be read as the amplitudes of all scattering events between $n$ ``high energy'' states and $m$ ``low energy'' states.
In all three panels, one can see that $H^4L^0$ has the largest fraction of large interaction elements, whereas $H^0L^4$ has the lowest,
in agreement with the conclusion that was drawn from Fig.~\ref{fig_Hop-Elem_flatten}.
We can also note that the interaction elements are significantly smaller for the strained system than the unstrained system.
It should also be quite clear from the lower panels that the elements do not follow a Gaussian distribution but are much more sharply peaked around $\left\langle J\right\rangle =0$. 

To summarize, low energy hopping elements are much smaller than high energy elements,
and elements in the strained system are smaller than in the unstrained.
This can be understood as follows:
The hopping terms $J$ depend on the $\Pi$ and $\Gamma$ in Eq.~\ref{eq_PI_and_Gamma}.
The observation we make is that in the strained system, the low energy states of $\mathcal{H}_{0}$ tend to have the $c^{A}$s living predominantly in the bulk,
whereas the $c^{B}$s mostly live on the boundary, see again Fig.~\ref{fig_StateStructure}.
This implies that one of $\Pi$ and $\Gamma$ will always be close to zero,
thereby suppressing the coefficient $J$ for low energy states.
However, for the states higher up in energy, the division between bulk and boundary is less well defined,
and this allows for $\Pi$ and $\Gamma$ to be nonzero at the same time, generally causing larger values of $J$.
Similarly, in the unstrained system, the sub-lattice division between bulk and boundary is not as strict, enabling the interaction elements to be larger.
We note that, beyond the observation of low energy states tending to have lower interaction coefficients, there is not much structure in the hopping matrix.

While this sounds like bad news for a sizeable coupling between low-energy states, there is also a silver lining.
The structure factor corresponding to the coupling between the states itself could be rather big in the strained system at low energies.
The reason for that is that the system is near degenerate, involving very small energy denominators in the perturbation theory.
Also, the degeneracies that come from the dangling boundary Majoranas depicted in Fig.~\ref{fig_BorderBonds} could act to boost the size of the structure factors.
We leave this question for future studies.

\subsection{Density of states of the interaction Hamiltonian}
In the previous section, we noted that the interaction elements $J_{ijkl}$ were much smaller than the bandwidth of the landau band.
For instance, for $\alpha=0.2$, the band edge is at $E/J\approx 0.2$, as seen in the inset of Fig.~\ref{fig_Hop-Elem_flatten}b).
At the same time, most of the interaction elements are $10^{-4}f$ or smaller.

However, the size of the interaction elements does not tell the whole story for a many-body Hamiltonian.
Since $\HamI$ in equation \eqref{eq_H_i_short} does not preserve any quantum numbers, the total number of nonzero terms on any row/column will scale as $N_A^2\times N_B^2$.
If all of the elements were added up coherently, one would expect an enhancement of $10^4$ already for $N_A=N_B=10$.
Even if we assume the elements are added up with random signs, simple statistics arguments will still give that we should expect eigenvalues to be a factor of $\sqrt{N_A^2N_B^2}=N_AN_B$ larger than the typical interaction element $J_{ijkl}$.

\begin{figure}[t]
  \begin{centering}
    \includegraphics[width=0.95\linewidth]{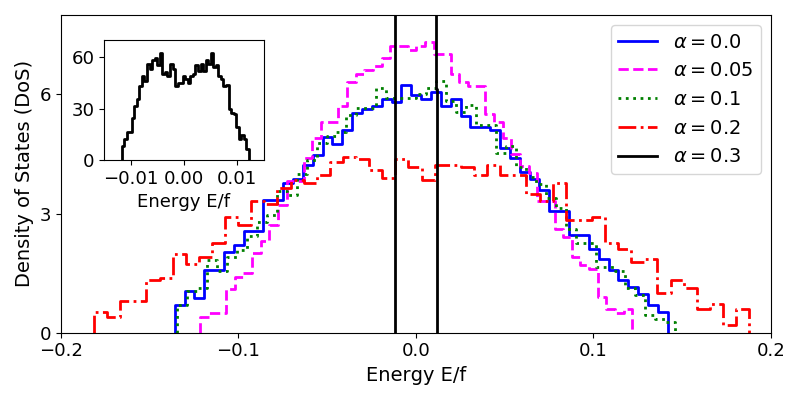}
    \par\end{centering}
    \caption{The many-body-bandwidth of the Majorana interaction, obtained by diagonalizing the Hamiltonian \eqref{eq_H_i_short} for $N_A=N_B=12$ Majoranas.
      Single particle energies are here (artificially) set to zero.
      The bandwidth is approximately $0.15 f$, independent of the value of $\alpha$; the exception being $\alpha=0.3$ where the width is $0.015 f$.
      For $f=J$, this width is comparable to the single-Majorana spectrum in the landau band.
    \label{fig_BandWidth}}
\end{figure}

The predicted enhancement is shown in Figure~\ref{fig_BandWidth}.
In the figure, we plot the density of states (DoS) obtained when diagonalizing \eqref{eq_H_i_short} for $N_A=N_B=12$ Majoranas, and $\alpha=0$, 0.05, 0.1, 0.2, 0.3.
Here, we only use the many-body part of the Hamiltonian and (artificially) set the single particle energies to zero.
Some more technical details are given in Section~\ref{sec:LevStat}, and Appendix.~\ref{app:LevStatTechnical}.

We see that the many-body bandwidth is approximately $0.15 f$, independent of the value of $\alpha$.
The only exception here is the case $\alpha=0.3$, where the bandwidth is much smaller ($0.015 f$), and also, the DoS shows signatures of a double peak.
The double peak is related to the fact that we choose to work with a binary $f$.
It arises if there are a few coefficients that are much larger than the bulk of the interaction elements and thus dominate the spectrum.
In the thermodynamic limit, we expect the double peak to smear out.

This bandwidth for $J=f$ is comparable to the bandwidth of the single-Majorana eigenstates, making it potentially a significant effect.

\subsection{Level statistics}\label{sec:LevStat}
In this section, we investigate whether the distribution of the couplings $J$ in the effective low-energy model is random enough to realize SYK type physics.
One way to see whether $\HamI$ 'potentially' realizes SYK physics is to study its level statistics.
Although level statistics is not a definitive test, it is an observable that is readily available through finite-size ED calculations.
The level statistics of the pure SYK model was studied in Refs.~\cite{You2017,Haque2019}.
The main result is that the random matrix classification depends on the number of Majoranas $N_{\chi}$ in the system.
In general, one expects to find that the level statistics fall into the classes of the Gaussian Unitary Ensemble (GUE), Gaussian Orthogonal Ensemble (GOE),
Gaussian Symplectic Ensemble (GSE), or the Poissonian Ensemble (P). 
To be precise, one should expect the classification of the SYK model to be cyclic modulo 8 and follow the pattern in the upper row of table \ref{tab:LevStat}.

\begin{table}[t]
  \begin{tabular}{c|c|c|c|c|c|c|c|c}
    $N_{\chi}$ (mod 8) & 0 & 1 & 2 & 3 & 4 & 5 & 6 & 7\tabularnewline
    \hline 
    $\HSYK$
    & O & O & U & S & S & S & U & O\tabularnewline
    $\HbSYK$
    & 2O & 2O & O & U & U & U & O & 2O\tabularnewline
  \end{tabular}
  \caption{Level statistics of the Majorana fermion SYK model,
$\HSYK$ as compared to $\HbSYK$.
    Here S=GSE, U=GUE, O=GOE and 2O = 2$\times$GOE are the different universal random matrix ensembles.
    Note how the level statistics of $\HbSYK$ traces those of $\HSYK$ but with a reduction in the symmetry classifications of one step such that S $\to$ U $\to$ O $\to$ 2O.}
  \label{tab:LevStat}
\end{table}

In a parallel publication, Ref.~\onlinecite{Fremling2021}, we show that the b-SYK model, described by $\HbSYK$ is distinct from the SYK model in that regard.
Due to its bipartite nature, it displays level statistics that is shifted as compared to the standard SYK-model.
For the b-SYK model, the level statistics has the same periodicity of 8,
but with the classification shifted as GSE$\to$GUE, GUE$\to$GOE, and GOE$\to$2$\times$GOE.
The last class, 2$\times$GOE, is obtained when two GOE spectra are superimposed and is distinct from the Poissonian spectral class. 

In order to determine the level statistics we choose to compute the gap ratio statistic, following \eg Refs.~\cite{Oganesyan2007,Atas2013}.
One starts with calculating the finite size spectrum $E_n$ which is ordered from lowest to highest energy.
From that one defines the set of level spacings $s_n = E_{n+1}-E_n$, from which the gap ratio 
\begin{equation}
r_n = \frac{{\rm min}(s_n, s_{n-1})}{{\rm max}(s_n, s_{n-1})}\;,\label{eq_gap_ratio}
\end{equation}
is obtained.
The statistics of the gap ratio $r_n$ has the advantage over the level spacing $s_n$ that it is automatically scaled to be in the range $0\leq r_n\leq1$ and no compensation for the local density of states is needed.

For Poisson statistics, the probability distribution of $r_n$ is $P(r)=2/(1+r)^2$ with mean value $\langle r \rangle=2\ln 2-1 \approx 0.39$.
For the Wigner-Dyson ensembles, the probability distributions are well-approximated by the surmise~\cite{Atas2013} $P(r) \propto (r+r^2)^{\beta}/(1+r+r^2)^{1+3\beta/2}$ up to normalization, with $\beta=1$ for GOE, $\beta=2$ for GUE, and $\beta=4$ for GSE.
Consequently, the averages are given by $\langle r \rangle_{\rm GOE}\approx 0.53$,
$\langle r \rangle_{\rm  GUE}\approx 0.60$,$\langle r \rangle_{\rm  GSE}\approx 0.67$.
For the $2\times$GOE distribution the average is $\langle r \rangle_{\rm  2\times GOE}\approx 0.42$\cite{Giraud2020,Fremling2021,Fremling2022}.

\begin{figure}[t]
  \begin{centering}
    \includegraphics[width=0.95\linewidth]{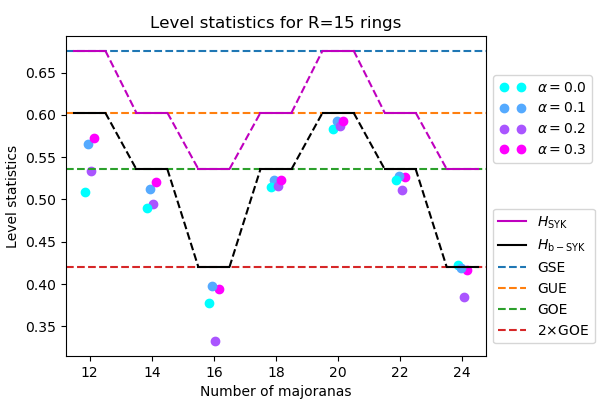}
    \par\end{centering}
    \caption{Level statistics as average gap-ratio $\rEXP$, for $R=15$ rings,
with strain $\alpha=0.0,0.1,0.2,0.3$ for $\HamI$ in \eqref{eq_H_i_short}.
We observe that the maximum $\rEXP$ for $\HamI$ is always lower than that of ideal SYK model ({\color{magenta}magenta} line),
irrespective of the amount of strain, but is in agreement with the b-SYK model (black line).
      \label{fig_LevStatStrainIII}}
\end{figure}

We numerically construct and diagonalize the many-body Hamiltonian in Eq.~\eqref{eq_H_i_short}, with $N_{\chi}$ Majoranas,
where the interaction elements are generated using the actual wave-functions.
The systems we consider consist of a fixed size of $R=15$ rings.
We generate one realization for the strain $\alpha=0.0,0.1,0.2,0.3$.
The results are presented in Fig.~\ref{fig_LevStatStrainIII}.

From Fig.~\ref{fig_LevStatStrainIII} we make the following observations:
as expected, the average gap-ratio of $\HamI$ is always lower than the average gap-ratio expected from the ideal SYK model.
Furthermore, the maximum average gap-ratio that is obtained (for any realization) of $\HamI$ follows the expected result for the b-SYK model.

Throughout this analysis, we make a simplifying assumption that we consider the kinetic energy to be quenched, \ie,
the Landau level bandwidth is assumed to be much smaller than the interaction strength. 
In the case with no strain this assumption is unjustified for obvious reasons.
In other words, we only used $\HamI$ and thus implicitly enforced the band broadening to be zero.
Thus, the results for systems with lower strain $\alpha<0.3$ only serve illustrational purposes and should be taken with a grain of salt.
Some further technical remarks regarding the calculation are given in Appendix.~\ref{app:LevStatTechnical}.

\section{Summary and Discussion}\label{sec:summary}
In this work, we have investigated whether there exists a route to realizing an SYK-type model
in the low energy limit of a Kitaev honeycomb model under strain.
We find a variant of the SYK model, called the b-SYK model, which has a bipartite structure
that is inherited from the bipartite structure of the original Kitaev honeycomb model. 
We argue that perturbations due to Heisenberg and $\gamma$-type couplings can, in principle, induce couplings between the Majorana fermions.
Assuming they have a random sign structure that can result from disorder in the system, these couplings show the structure required to realize SYK-type physics.
We show this by analyzing the level statistics of the low energy Hamiltonian, $\HamI$.
While we cannot give a definitive answer to whether we believe the model is realized in more realistic setups, we find evidence that warrants an even more in-depth analysis of this question.

There are several open questions.
Firstly, the strength of the interaction matrix elements appears to be very low.
This is mostly due to the spatially separated wave-functions of the $A$- and $B$-Majorana fermions in the strained system. Potentially,
this could be overcome due to a large structure factor emerging in perturbation theory within the near-degenerate low-energy states. Secondly,
but related, we discarded the kinetic energy of the model completely.
This is an approximation, and its validity also hinges on the strength of the effective coupling constants.
Third, the random structure should be verified in a more realistic setting, including disorder. 

In order to make further progress, we believe that one must switch to more sophisticated numerical methods.
A two-dimensional version of DMRG would allow to implement the Heisenberg and ``cross'' term couplings directly exactly and then deal with the low energy limit of the strongly interacting microscopic model. Alternatively,
one could attempt to use high-order series expansions to get a better estimate of the effective interactions between the Majorana fermions.

Finally, the presence of next-nearest neighbor couplings could lead to an actual SYK model since it couples Majoranas on the same sublattice. In that case, one would not expect the stark suppression of matrix elements due to the spatial localization, and lower order corrections could also potentially be relevant.

Lately, a related work~\cite{Agarwala2020} investigated the fate of generic perturbations in the strained Kitaev honeycomb model.
In their work, translational invariance was assumed in the low-energy limit. In this work, we explicitly model finite-size systems, 
so some of the features discussed here cannot show up in their setup.
One example is the coupling between $A$- and $B$-Majorana fermions.
The reason for that is that in the continuum model, the boundary is pushed all the way to infinity.
Consequently, instead of an SYK-like model, they find intricate behavior akin to fractional quantum Hall physics, and it remains an open and interesting question how the two works relate to each other and connect for finite system sizes.

\section*{Acknowledgements}
We would like to thank Maria Hermanns, Graham Kells, Matthias Vojta, Stephan Rachel, Philippe Corboz and Masudul Haque for useful discussions and work on related problems.
This work is part of the D-ITP consortium, a program of the Netherlands Organisation for Scientific Research (NWO) that is funded by the Dutch Ministry of Education, Culture and Science (OCW).

\bibliography{KitaevBib}

\appendix

\section{Counting degrees of freedom on a finite open lattice}
\label{app:CountingDOFs}
Let us explicitly count the degrees of freedom for an open lattice.
We assume that there are $N$ sites and, for simplicity, that $N$ is \emph{even} such that there is an \emph{even} number of $\hat{c}_{i}$ operators.
The case with \emph{odd} number of sites can be handled, too, but one needs
to be a bit more careful with the presence of a dangling $b$-Majorana fermion.
In the spin language, there are $N$ sites, which amounts to $2^{N}$ degrees of freedom.
When expanding into $4N$ Majoranas, one would naively have $2^{2N}$ degrees of freedom
(since each pair of Majoranas is one degree of freedom).
Formally, with the onside projectors $D_{j}$, we introduce the $2^{N}$ local restrictions necessary to bring the number down to $2^{N}$ again.

How do we now obtain the same counting using the bond variables $u_{i,j}$?
Let us compute the maximum number of possible bond variables and flux plaquettes for a generic lattice.
Each site contributes $3$ $b$-Majoranas for a total of $3N$ $b$-Majoranas.
These $3N$ $b$-Majoranas will form at most $\frac{3}{2}N$ bonds
(this is why $N=$even is important).

In the Hamiltonian~\eqref{eq_KM_Defintion} there will be bonds along the boundary that are not present, and thus effectively are zero.
For the sake of argument, we say that any bond not in the Hamiltonian has a coefficient $\epsilon\to0$, but formally still does exist.
For $H_K$ this means that the effect of the missing boundary bonds is that Eq.~\eqref{eq_KM_Defintion} has a degeneracy that grows exponentially with the length of the boundary.

\begin{figure}
  \begin{centering}
    \includegraphics[width=0.60\columnwidth]{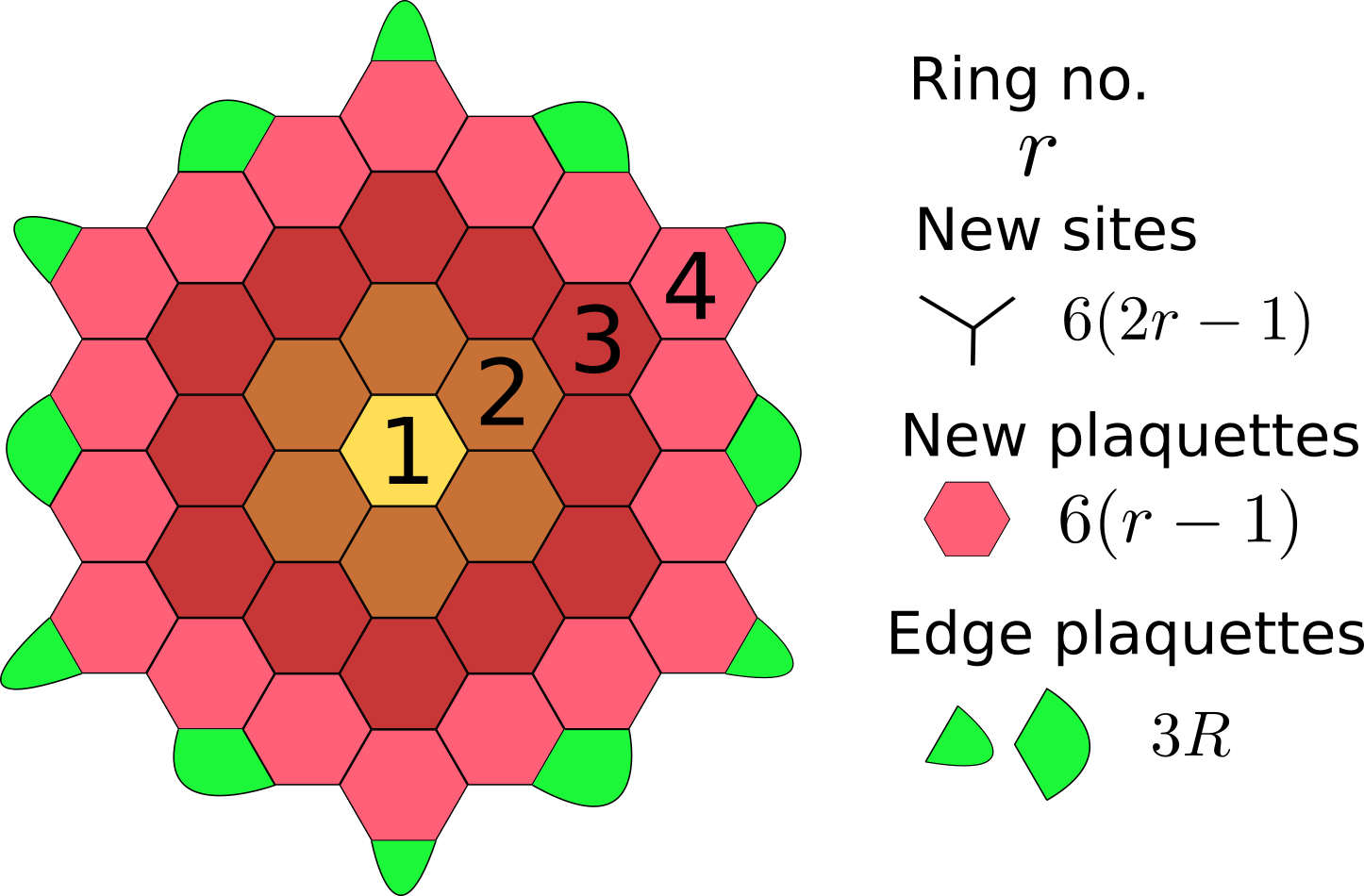}
\par\end{centering}
\caption{Sketch of the counting of plaquettes and sites for the Honeycomb lattice. At row $r$,
$6(2r-1)$ sites forming $6(r-1)$ plaquettes are added, and there are $3R$ fictitious plaquettes on the boundary.
  \label{fig_Sketch_plaquette}}
\end{figure}

The number of flux plaquettes is a bit trickier.
Let's see how this works for the open Honeycomb lattice.
For $R=1$ there is only one plaquette with $N=6$ sites, while for every row $r>1$ then $6(r-1)$ plaquettes and $6(2r-1)$ sites are added to the honeycomb.
On the edge of the honeycomb there are also $6R$ dangling bond Majoranas forming $3R$ fictitious edge plaquettes. 
The total number of sites is $N=6\sum_{r=1}^{R}(2r-1)=6R^2$,
while the number of plaquettes is $N_{\lambda}=1+6\sum_{r=2}^{R}(r-1)+3R=1+3R^2=\frac{N}2+1$.
A sketch of this counting can be found in Fig.~\ref{fig_Sketch_plaquette}.

Thus, we have $\frac{N}{2}+1$ flux sectors and $N$ $c$-Majoranas.
Since the $c$-Majoranas contribute $\frac{N}{2}$ degrees of freedom,
naively it looks like we have $\frac{N}{2}+\left(\frac{N}{2}+1\right)=N+1$ degrees of freedom.
This is of course \emph{one} too many.
The last constraint comes from the site projectors $D_{i}$.
Simply put, one can argue that $D=\prod_{i}D_{i}$ can be rewritten as 
\[
D\propto\overbrace{\prod_{\left\langle i<j\right\rangle }\hat{u}_{ij}}^{U}\times\overbrace{\prod_{i}c_{i}}^{C}
\]
such that $D=UC$ is a constant.
This constraint, in conjunction with $Dc_{i}=-c_{i}D$, leads to the constraint that the physical space either has an even or odd number of fermions.
Thus, from our set of $\frac{N}{2}$ Majoranas giving a Hilbert space of $2^{\frac{N}{2}}$ we remove half
of the states giving $2^{\frac{N}{2}-1}$ states and arrive at $\left(\frac{N}{2}+1\right)+\left(\frac{N}{2}-1\right)=N$
degrees of freedom. 

We now argue that we can deform the above-described honeycomb lattice into any other trivalent lattice,
by pairwise swapping of bonds without changing the number of independent plaquettes.
For our purposes, this argument then applies to cases where there are \eg missing sites due to disorder.

\begin{figure}
  \begin{centering}
    \includegraphics[width=0.6\columnwidth]{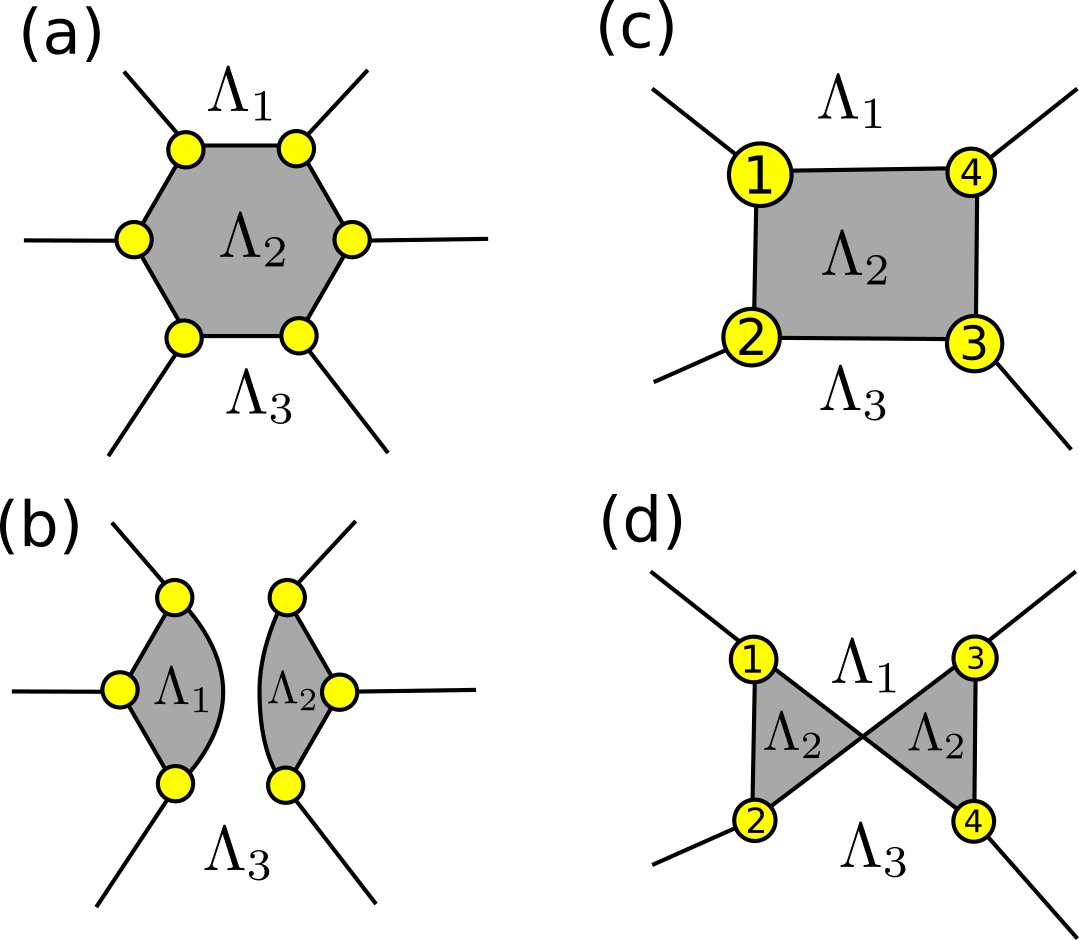}
\par\end{centering}
\caption{(a)-(b) Swapping of bonds in a planar diagram that manifestly preserves the number of bonds.
  (c)-(d) Swapping of bonds leading to a non-planar diagram, which at fist glance has one extra plaquette.
However bond-crossings should be counted as a ``-1'' for the purposes of plaquettes.
  \label{fig_Swap}}
\end{figure}

For planar lattices, the argument is straight forward, and an example of a plaquette preserving swap is depicted in Fig.~\ref{fig_Swap}(a) and (b).
For a non-planar graph, the same arguments still hold but can be harder to visualize.
The key is that bonds that cross in a non-planar diagram add a -1 to the number of plaquettes.
See Fig.~\ref{fig_Swap}(c) and (d) for an example.

\section{The Majorana SVD}\label{app:Majoran-svd}
In a disordered system, it is not guaranteed that the number of $A$
sites, $N_{A}$, and the number of $B$ sites, $N_{B}$, are the same.
Let us look a bit more carefully at the case when $N_{A}\neq N_{B}$.
Without loss of generality, we assume that $N_{A}<N_{B}$.
Restoring the index counting, the Hamiltonian in Eq.~\eqref{eq_H_k_bipartite} then reads
\begin{align*}
H_K & =\i\sum_{i=1}^{N_{A}}\sum_{j=1}^{N_{B}}c_{i}^{A}C_{ij}c_{j}^{B}=\i\sum_{n=1}^{N_{\gamma}}S_{n}a_{n}^{A}a_{n}^{B}
\end{align*}
 where $N_{\gamma}=\min\left(N_{A},N_{B}\right)=N_{A}$ (This is known
as the compact svd). The transformation between the $c$ and $a$
Majoranas then reads

\begin{align}
a_{n}^{A} & =\sum_{i=1}^{N_{A}}c_{i}^{A}U_{in}\nonumber \\
a_{n}^{B} & =\sum_{j=1}^{N_{B}}V_{nj}^{T}c_{j}^{B}\;.\label{eq_Maj_Map}
\end{align}
For the compact SVD it holds that $U^{T}U=V^{T}V=1_{N_{\gamma}}$.
Thus we have
\[
\sum_{i=1}^{N_{A}}U_{n^{\prime}i}^{T}U_{in}=\delta_{n^{\prime}n}^{\left(N_{\gamma}\right)}=\sum_{j=1}^{N_{B}}V_{n^{\prime}j}^{T}V_{jn}\;.
\]
 However, it is only true that $\sum_{n=1}^{N_{\gamma}}U_{in}U_{ni^{\prime}}^{T}=\delta_{i,i^{\prime}}^{\left(N_{A}\right)}$
and not $\sum_{n=1}^{N_{\gamma}}V_{jn}V_{nj^{\prime}}^{T}\neq\delta_{j,j^{\prime}}^{\left(N_{B}\right)}$.
From this we see that in this representation we cannot invert equation
Eq.~\eqref{eq_Maj_Map} to obtain $c_{j}^{B}$ as a function of $c_{n}^{B}$.

For this to work, we actually need to consider the regular SVD, where
\begin{align*}
C_{ij} & =\sum_{n=1}^{N_{A}}\sum_{m=1}^{N_{B}}U_{in}S_{n}\delta_{n,m}^{\left(N_{A},N_{B}\right)}V_{mj}^{T}\\
 & =\sum_{n=1}^{N_{A}}\sum_{m=1}^{N_{A}}U_{in}S_{n}V_{nj}^{T}\;.
\end{align*}
 In this representation $U$ is a $N_{A}\times N_{A}$ unitary and
$V^{T}$ is a $N_{B}\times N_{B}$ unitary. Note that in the compact
form the last $N_{\Delta}=N_{B}-N_{A}$ rows of $V^{T}$ are dropped.
This means that there are $N_{\Delta}=N_{B}-N_{A}$ number of $c_{n}^{B}$
Majoranas that are not present in the Hamiltonian and thus allow for
degeneracies.
Thus, in order to invert Eq.~\eqref{eq_Maj_Map} we need
the extended matrices, such that 

\begin{align*}
c_{i}^{A} & =\sum_{n=1}^{N_{A}}a_{n}^{A}U_{ni}^{T}\\
c_{j}^{B} & =\sum_{m=1}^{N_{B}}V_{jm}a_{m}^{B}\;.
\end{align*}

\section{Technical remarks on Level statistics and Majorana diagonalization}\label{app:LevStatTechnical}
In this section, we summarize some technical remarks regarding the level statistics calculation in Section~\ref{sec:LevStat}.
The number of Majoranas is only conserved modulo 2, and the Hilbert space can be split into many-body states with either an even or odd number of Majoranas.
Fortunately, this is also the Hilbert space constraint that the Kitaev construction puts on our Majoranas, \ie,
that the physical space has only many-body states with an even or odd number of Majoranas.
The party constraints of the two models are thus commensurate.

We also note that with $N_{\chi}$ Majoranas we can form up to $N_{f}=\left\lfloor \frac{N_{\chi}}{2}\right\rfloor$ fermions by enforcing that say $a_{j}\ket 0=\i a_{N_{\chi}-j+1}\ket 0$ for $j\leq N_{f}$.
This is a relation that comes automatically from solving $H_K$ as it has $\i a_{\lambda}^{A}\ket 0=a_{\lambda}^{B}\ket 0$.

Sometimes, such as when the level statistics is GSE type, there are exact degeneracies.
These degeneracies must be handled by pruning the spectrum to get rid of this residual symmetry,
before Eq.~\eqref{eq_gap_ratio} in the main text can be applied. 

Further, to flush out the level statistics, one needs to average over many realizations. One would like at least a thousand energy level samples to resolve the level statistics satisfactorily.
In this work we will \emph{not} average over several $f_{j,kl}$-realizations, since computing all the $J_{ijkl}$'s from Eg.~\eqref{eq_j_of_pi} is numerically costly.
Instead, we will fix the $f$-realization and average over several sets of Majoranas from the same realization.
Let us comment on the ``averaging'' that is applied here.
From Fig.~\ref{fig_Hop-Elem_flatten} and Fig.~\ref{fig_ElemStats} we know that certain Majoranas are almost decoupled from the other Majoranas in terms of the sizes of interaction elements.
As a result, these Majoranas will affect the spectrum in a similar way that a symmetry will affect the spectrum and thus pull the average gap ratio towards smaller values.
To avoid this, we randomly sample the $N_\chi = 2\cdot N_f$ Majoranas only from the set of  $\tilde N$ Majoranas with the largest $\sqrt{w^Aw^B}$.
We choose $\tilde N$ as the smallest number where ${{\tilde N}\choose{N_f}} > 10\sqrt{N_{\mathrm{tries}}}N_{\mathrm{tries}}$,
where the factor $10\sqrt{N_{\mathrm{tries}}}$ is chosen to reduce the risk of sampling the same Majoranas twice. 
In order to obtain good statistics we repeat the sampling (from the $f$-realization) $N_{\mathrm{tries}}$ times.

\end{document}